\documentclass{jaa}
\usepackage[usenames,dvipsnames]{color}
\usepackage{natbib}
\usepackage[hyperfootnotes=false]{hyperref}

\hypersetup{colorlinks=true, citecolor=blue, linkcolor=blue,  filecolor=blue, urlcolor=blue}
\usepackage[english]{babel}
\usepackage{epsfig}
\usepackage{epstopdf}
\usepackage{psfrag}
\usepackage{color}
\usepackage{graphicx}
\usepackage{fancyhdr}
\usepackage{graphics} 
\usepackage{soul}

\usepackage{amsmath}
\usepackage{xspace}
\usepackage{xcolor}

\newcommand{\dtec}{$\delta\rm{TEC}$\xspace}
\newcommand{\hi}{H{\sc i}\xspace}
\begin{document}\sloppy
%%paper title
%%For line breaks \\ can be used within title
%\title{Effects of Earth's ionosphere on sensitive radio interferometers}
\title{Exploring Earth's Ionosphere and its effect on low radio frequency\\observation with the uGMRT and the SKA}
%%author names are separated by comma (,)
%%use \and before the last author name
%%use a * along with the number separated by comma
%% for the  author for correspondence
%%\textsuperscript{number} is used for affiliation
%%\affilOne, \affilTwo etc., upto \affilTwentyfive is possible
%%Please note the first letter after \affil is capitalised in the command
%%
%
\author{Sarvesh Mangla\textsuperscript{1,*}, Sumanjit Chakraborty\textsuperscript{1,2}, Abhirup Datta\textsuperscript{1} and Ashik Paul\textsuperscript{3} }
\affilOne{\textsuperscript{1}Department of Astronomy, Astrophysics and Space Engineering, Indian Institute of Technology, Indore 453552, Madhya Pradesh, India.\\}
\affilTwo{\textsuperscript{2}Space and Atmospheric Sciences Division, Physical Research Laboratory, Ahmedabad 380009, Gujarat, India.\\}
\affilThree{\textsuperscript{3}Institute of Radio Physics and Electronics, University of Calcutta, Kolkata 700009, West Bengal, India.\\}
%
%%escape two column mode for title, affiliation and abstract
%%by giving \twocolumn command as shown
%
\twocolumn[{
\maketitle
%
%%include \corres to print the corresponding author Email id
\corres{mangla.sarvesh@gmail.com}

%%include \msinfo for
%%manuscript information such as
%%received, revised and accepted dates
%%
% \msinfo{1 January 2015}{1 January 2015}

%%abstract
\begin{abstract}
The Earth's ionosphere introduces systematic effects that limit the performance of a radio interferometer at low frequencies ($\lesssim 1$\,GHz). These effects become more pronounced for severe geomagnetic activities or observations involving longer baselines of the interferometer.  The uGMRT, a pathfinder for the Square Kilometre Array (SKA), is located in between the northern crest of the Equatorial Ionisation Anomaly (EIA) and the magnetic equator. Hence, this telescope is more prone to severe ionospheric conditions and is a unique radio interferometer for studying the ionosphere. Here, we present 235\,MHz observations with the GMRT, showing significant ionospheric activities over a solar minimum. In this work, we have characterised the ionospheric disturbances observed with the GMRT and compared them with ionospheric studies and observations with other telescopes like the VLA, MWA and LOFAR situated at different magnetic latitudes. We have estimated the ionospheric total electron content (TEC) gradient over the full GMRT array which shows an order of magnitude higher sensitivity compared to the Global Navigation Satellite System (GNSS). Furthermore, this article uses the ionospheric characteristics estimated from the observations with uGMRT, VLA, LOFAR and MWA to forecast the effects on the low-frequency observations with the SKA1-MID and SKA1-LOW in future.
\end{abstract}
%%insert keywords separated by 3 hyphens using \keywords{words}
}]
%
%%close the twocolumn escape here
%
%%include \doinum{number}for the DOI number in the header
%%include \volnum{number} for the volume number in the header
%%include \year{yyyy} for  year of publication in the header
%%include \pgrange{num--num} page range of article in the header
%%include \artcitid{num} for the article citation id
%%include \lp to print last page of the article
%%include \setcounter{page}{pagenum} for the exact starting page of the article

% \doinum{12.3456/s78910-011-012-3}
% \artcitid{\#\#\#\#}
% \volnum{000}
% \year{0000}
% \pgrange{1--}
% \setcounter{page}{1}
% \lp{1}

\section{Introduction}
\label{sec:introduction}
The impact of the Earth's ionosphere is one of the major challenges in low-frequency ($\lesssim$\,1.0\,GHz) radio observations. It is a perturbing medium for transionospheric radio signals measured by Earth-based radio telescopes, which suffer from phase corruption. To perform sensitive measurements at low radio frequencies, it is necessary to accurately calibrate the ionosphere-induced path length differences from the measured signal. \par
The ionosphere consists of partially ionized plasma, which extends from about 60\,km to beyond 1000\,km altitude, transitioning smoothly into the plasmasphere. The ionosphere is formed as X-ray, and extreme ultraviolet (EUV) light from the Sun provides the energy that ionizes the atoms and molecules present in the upper atmosphere. At nighttime, the recombination process occurs, and the number of electron-ion pairs decrease. Other sources such as cosmic rays, can ionize the atmosphere, though not nearly as strong as the Sun. At night, the electron density peak lies at an altitude of about 250-500\,km \citep[see][]{Mannucci1998RaSc...33..565M}. The ionosphere is a highly dynamic region that varies with latitude and time. The electron density changes dramatically from local daytime to nighttime, peaking in the early afternoon and progressively declining after midnight. The electron density increases as one descends in the latitude of both hemispheres, reaching a maximum around the geomagnetic equator with a trough somewhere between the geographic and geomagnetic equators. As a result, the ionospheric activity is continuously varied during the daytime and has unanticipated changes during the nighttime, especially at low-latitude regions ($\pm$20\,degree magnetic latitudes). \par
The ionosphere has been investigated for decades with ionosondes and global navigation satellite system (GNSS) at a rough resolution (temporally 2\,hour, spatially $5^{\circ}$ and $2.5^{\circ}$ in longitude and latitude, respectively; \citep[see][]{arora2015PASA...32...29A}. The ionosphere is typically measured in terms of column density of electrons along the line of sight (LoS) or the total electron content (TEC) where a TEC unit or 1\,TECU=$\rm 10^{16}$ electrons/$\rm m^2$. These electrons act to refract the incoming transionospheric signals of far-field astronomical sources, introducing a delay that is often observed as a phase change in the radio observation. Because an interferometer measures the phase difference across baselines, it cannot detect the total amount of phase contributed because of the ionosphere. Instead, an interferometer measures the \dtec caused by the spatial variations in the ionosphere. \par 
The Giant Metrewave Radio Telescope (GMRT) \citep[][]{swarup_GMRT} is one of the largest fully operational sensitive telescopes at low frequencies ($\lesssim$\,1\,GHz). The configuration and geographic location (latitude = $19^{\circ} \, 05' \, 35.2''$ N and longitude = $74^{\circ} \, 03' \, 01.7''$ E) makes this interferometer uniquely sensitive to ionospheric disturbances between the magnetic equator and the northern crest of the Equatorial Ionization Anomaly (EIA) \citep[see][and references therein]{Appl1946Natur.157..691A,sc37,sc39,Deepthi2020AdSpR..65.1544A,sc41} in the Indian longitude sector. GMRT consists of 30 dishes (each of 45\,m diameter), 14 of them are randomly placed in the central square of $\rm 1.4\times1.4\,km^2$ area, while the remaining 16 dishes are placed along three arms, each of 14\,km (approximately) in a `Y' shaped configuration. Recently, GMRT has been upgraded to uGMRT \citep[][]{Yashwant2017CSci..113..707G}, which has added extra capability in terms of frequency coverage and sensitivity. Some of the main features are (1) wide frequency coverage, from 120 to 1450\,MHz subdivided into four bands, replacing the original five bandwidth frequency bands centered around 150, 235, 325, 610, 1420\,MHz; (2) Instead of the 32\,MHz bandwidth of the original GMRT design, the maximum bandwidth is now 400\,MHz; (3) Improved receiver systems and better dynamic range\footnote{The ratio between the peak flux on the image and root mean square noise in a region believed to be source free region.}; (4) The upgradation has been implemented with the least possible disruption to the availability of the existing GMRT data for scientific observations. The configuration of uGMRT and wide frequency coverage gives us the ability to observe and study the ionospheric fluctuations over a broad range of scales discussed in \citet{Lonsdale2005ASPC..345..399L} simultaneously. The calibration of ionospheric effects is a challenge for any radio interferometer. Extensive studies with uGMRT may provide substantial information to calibrate out the ionospheric corruption from the signal for SKA1-LOW. \par
The upcoming SKA telescope \citep[see][]{Dewdney2009IEEEP..97.1482D, Braun2019arXiv191212699B} with exceptional sensitivity and huge collecting area will address a wide range of questions in astrophysics and cosmology. The SKA phase 1 (SKA1\footnote{The readers are referred to Project summary of the SKA1 \url{https://www.skatelescope.org/wp-content/uploads/2021/02/22380_SKA_Project-Summary_v4_single-pages.pdf}}) have started building at its two sites - South Africa and Australia. The SKA1-LOW telescope is being built in Western Australia at the Murchison Radio-astronomy Observatory (MRO). It will feature 512 ``field stations'' located in a central core along with three spiral arms stretching approximately 65\,km, with 256 dipole antennas in each station. This will be a low-frequency array operating from 50 to 350\,MHz. On the other side, SKA1-MID is being built in the Karoo region of the Northern Cape of South Africa. It will consist of 133 SKA dishes (each of 15\,m diameter) and an additional 64 dishes (each of 13.5\,m diameter) from the MeerKAT telescope, totaling 197 dishes of mixed array extending in a three-arm spiral configuration with a maximum baseline of 150\,km. This array will be operational in four different bands 0.35-1.05\,GHz, 0.95-1.76\,GHz, 4.60-8.50\,GHz, and 8.30-15.30\,GHz. As these new telescope facilities become operational with huge baselines and wide bandwidth, the ionosphere would corrupt radio interferometric observations, especially for SKA1-LOW observation and SKA1-MID band\,1, and performing calibration will become a challenging task. \par
The present paper is outlined as follows: in  Sec. \ref{sec:radars_ionosondes} we summarise the ionospheric studies conducted using Radars, Ionosondes, and GNSS. In Sec. \ref{sec:radio_interferometer_ionosphere} we present the ionospheric effect on electromagnetic radiation at low radio frequencies, as well as how the ionosphere affects the data collected from any radio telescope. In Sec. \ref{sec:antenna_based_method} we explore ionospheric structures with antenna-based method using 235\,MHz observation of GMRT. In Sec. \ref{sec:field_based_method} we present field-based method to study the ionospheric structures. In Sec. \ref{sec:scintillations} we review the results related to the scintillations obtained using GMRT. Further, in Sec. \ref{sec:future_SKA} we forecast the calibration challenges for future instruments like SKA because of ionospheric effects. Finally, we summarise this article in Sec. \ref{sec:summary}
\section{Ionospheric studies using Radars, Ionosondes and GNSS}
\label{sec:radars_ionosondes}
There has been substantial progress with the observational analysis of the ionospheric F region, and the scattering of signals from the irregularities in this region is obtained as spread F on spatial charts of radar data and ionograms \citep[see][]{sc02,sc01,sc03}. Initial signatures of plasma depletion were observed by the OGO-6, a satellite that is placed in the polar orbit \citep[see][]{sc04}. Ionospheric Scintillation (defined as random, intense, and rapid fluctuations (usually of the order of tens of seconds to few minutes; \citealt{Kintnerhttps://doi.org/10.1029/2006SW000260}) in the amplitude and phase of a transionospheric signal) study is crucial for understanding the spatial and temporal distribution of ionospheric irregularities and understanding the physical processes that lead to their formation. The ionospheric irregularities emerge as intense bite-outs in the in-situ density maps and cause scintillations in transionospheric satellite links \citep[see][]{sc05,sc06,sc07,sc08}. \par
Generally, the nighttime spread F irregularities occur with various scale sizes (from less than a meter to several kilometers), and the initial observations were reported by \citet{sc09}. Sensitive measurements using instruments like ionosondes, radars (coherent and incoherent scatter, HF Doppler), radiosondes, GNSS satellite receivers, and various in-situ measurements reveal the morphological features of spread F and its long-term characteristics such as its temporal, spatial, and seasonal dependence of geographic and geomagnetic factors. All these factors make the understanding of spread F phenomenon quite complicated in terms of prediction of space weather modeling \citep[see][]{sc10,sc11,sc12,sc13,sc14,sc15,sc16,sc17,sc18,sc19,sc20,sc21,sc03,sc22}. \par
The variability of decorrelation time, cumulative distribution functions of signal amplitude, fade duration, the distribution of phase and intensity rates, and depolarization effects caused by diffractive scattering have all been used to study the effects of scintillation on communication systems using data from the equatorial anomaly region \citep[see][]{sc23,sc24,sc25,sc26}. Even though the concepts of amplitude and phase scintillation are well developed \citep[see][and references therein]{sc27}, research efforts on phase scintillation are less extensive than those on amplitude scintillation \citep[see][]{sc28}. Given the same studies of phase information of the received signals from GNSS, especially at the sensitivity of 50\,Hz sampling rate, provides a glimmer of hope for characterizing phase scintillations in terms of frequency and intensity. \par
A recent development to study the ionosphere using the radar is the Ionospheric Field Station of the University of Calcutta, located at Haringhata (22.93$^\circ$N, 88.50$^\circ$E geographic; magnetic dip: 36.20$^\circ$N) about 50\,km North-East of Calcutta, that hosts a VHF radar operational at 53\,MHz. It is to be noted that this station is not only on the Tropic of Cancer but also is near the northern crest of the EIA. Some initial experiments with this VHF radar brought forward that the Field-aligned Irregularity (FAI) echoes at this location are similar to those observed at the off-equatorial low and middle latitude sites \citep[see][]{sc29,sc30,sc31,sc32,sc33,sc34,sc35,sc36}. As a result of the fact that the FAIs get generated by the gradient drift instability, some of these observed features could be linked with the sporadic E formed by tidal winds and gravity waves. Furthermore, as this location is situated in the deep convective zone that launches short-period gravity waves, these echoes, which are patchy in nature, are important observations. These gravity waves, accompanied by the associated winds, propagating up to the E-region of the ionosphere, could form sporadic E patches, which in turn becomes unstable and generates these FAIs. \par
While this section discusses the ``traditional" ionospheric probes, in the following section we describe a more sensitive probe, namely radio interferometers. \par
\section{Effect of the ionosphere on radio interferometers}
\label{sec:radio_interferometer_ionosphere}
A radio interferometer directly measures the ``spatial coherence function'' with the pair of antennas pointed towards the sky \citep[see][]{Cornwell1999ASPC..180..187C}. The function describes the far-field radiation pattern and by performing the Fourier inversion of this function, one gets an image of the intensity distribution of the sky signal. The observed visibility ($V_{ij}$) for a baseline ($i,j$) is given by the van-Cittert-Zernike theorem:
\begin{equation}
    \label{eq:visibility}
    V_{ij}(u,v) = G_{ij} \int  e^{-i (\phi_i(l,m) - \phi_j(l,m))}I(l,m) e^{-2 \pi i (u_{ij}l+v_{ij}m)} dl dm
\end{equation}
where $I(l,m)$ is the brightness distribution of the sky, $G_{ij}(t)$ is ``instrumental gain''. For an antenna $i$, the phase for a point source at sky position ($l,m$) expressed as $\phi_i(l,m)$, which is the effect of the ionosphere by simply adding a excess path length. However, the scale size of ionospheric disturbances is $\sim$ a few hundred kilometers, thus for a small field of view (FoV), it is often the case where $\phi_i(l,m)$ is constant over the entire primary beam. \par
\begin{figure*}[ht]
    \centering
    \includegraphics[width=1.5\columnwidth]{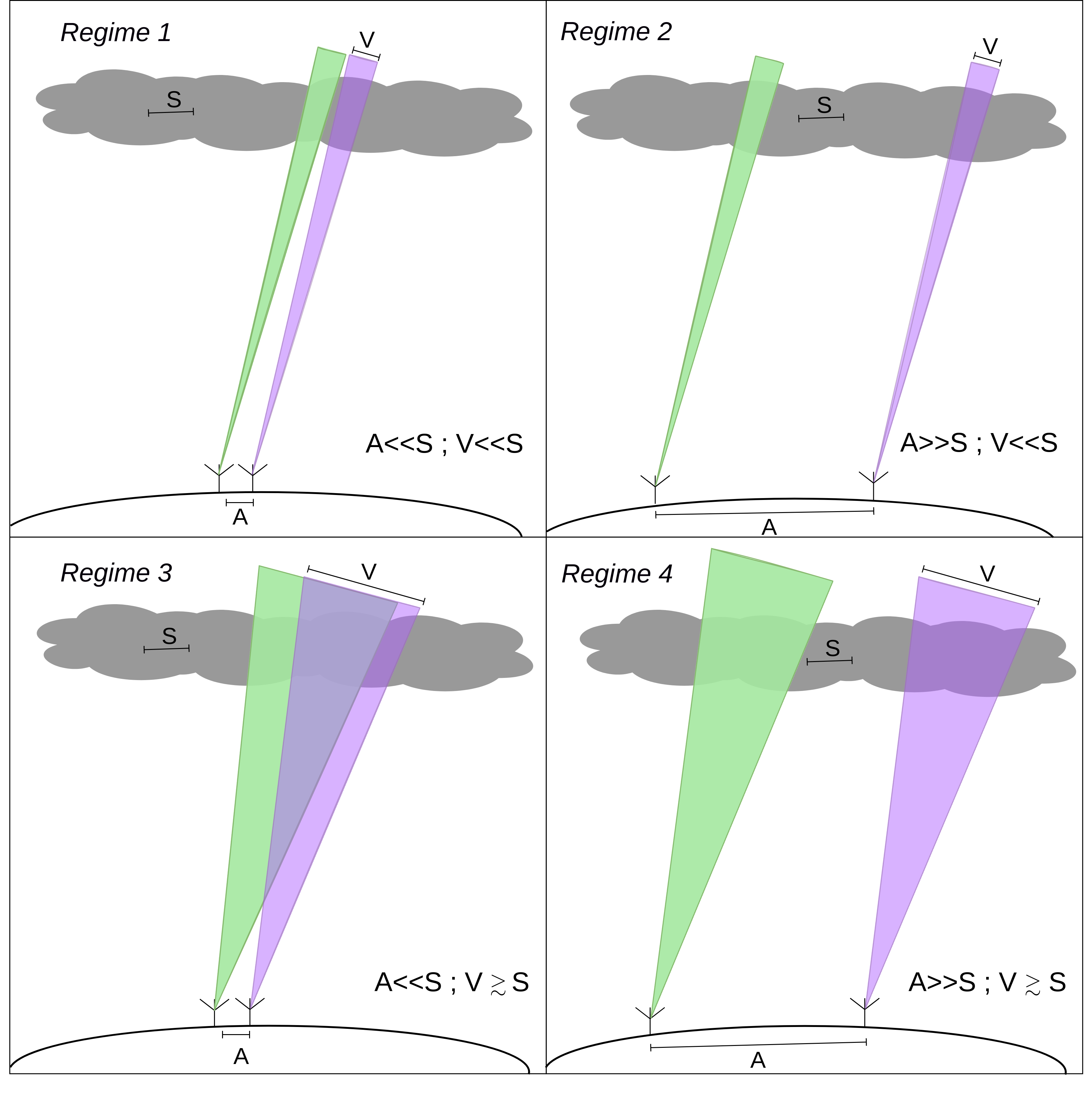}
    \caption{Schematic overview of four calibration regimes discussed by \citet{Lonsdale2005ASPC..345..399L} for low-frequency arrays. The quantities A, S, and V are the array size on the ground, the ionospheric irregularities scale size, and the FoV at projected ionospheric altitude respectively. For isoplanatic conditions (V$<<$S; Regime 1 and 2), the ionospheric phase rotation does not vary much within the FoV of each antenna. For non-isoplanatic conditions (V$\gtrsim$S; Regime 3 and 4), the ionospheric phase rotation per antenna varies over the FoV.}
    \label{fig:four_regimes}
\end{figure*}
Determining the effects of the ionosphere on low-frequency arrays is challenging. Calibrating the data for a continuously varying ionosphere is more challenging than calibrating for slowly varying instrumental gain. Any successful calibration technique must limit the parameters of the ionosphere above the array to the point where the interferometer phase is a delay on any baseline, towards any point in the field of view (FoV), at any moment during the observation, can be properly measured and corrected for. Based on the relationship between the three length scales which are the physical scale size of ionospheric fluctuations (S) with large enough electron content to cause considerable phase delay, different scales of the array (A) on the ground and the projected size of the FoV (V) of the individual antenna at a typical ionospheric height, \citet{Lonsdale2005ASPC..345..399L} discussed four calibration regimes for ionospheric phase calibration using these three length scales. In Fig.\ \ref{fig:four_regimes}, all four calibration regimes are shown. Here, the array is represented by two antennas on the ground observing through the ionosphere (grey structures) with the respective FoV (green, violet areas). \par
Due to relatively narrow beam patterns in regime 1 (A $<<$ S ; V $<<$ S) and regime 2 (A $>>$ S ; V $<<$ S), each antenna observes an approximately `constant' TEC across the FoV. Both these regimes, featuring a narrow FoV, are readily dealt with ``self-calibration''. Short enough time scale is sufficient to remove the ionospheric phase rotations from the visibilities. Relatively wide beam patterns in regime 3 (A $<<$ S ; V $\gtrsim$ S) and regime 4 (A $>>$ S ; V $\gtrsim$ S) cause the antenna to observe TEC `variation' across the FoV. For regime 3, TEC variation across the array for a single viewing direction within the FoV is approximately a gradient. This causes the apparent position of sources to change with time and viewing direction, but there is no source deformation. The calibration for such a compact array requires a minimal number of parameters because each antenna observes the same part of the FoV. For regime 4, TEC variation differs significantly from a gradient for a single viewing direction across the array. Individual lines of sight from different antennas to one source may trace completely different paths through the ionosphere. This causes a shift in the source position and can deform the shape, which varies with time and viewing direction. The calibration for such an extended array requires many parameters to compute a full 3-dimensional ionospheric phase model. \par
For radio interferometers, propagation delay is the dominant term, which is the effect of varying refractive index `n' of the ionospheric plasma along the wave trajectories. The total propagation delay integrated along the LoS at frequency $\nu$ results in phase rotation given by:
\begin{equation}
    \label{eq:phase_rotation}
    \phi_{\rm ion} = - \frac{2 \pi \nu}{c} \int (n - 1)\,dx
\end{equation}
For a constant `n' in space and time, a constant phase error will be present, resulting in a constant spatial shift of the observed sources compared to the true sky. However, things get more complicated when `n' strongly depends on space and time. \par
Furthermore, the refractive index `n' can be calculated exactly for a cold, collisionless, magnetized plasma \citep[see][]{davies1990ionospheric_book}. Signals with frequencies $\nu >> \nu_{p}$ (Plasma frequency; $\rm\sim\,10\,MHz$ for ionosphere), `n' can be expanded \citep[see][]{Datta2008RaSc...43.5010D_higherorderionosphericerror} using third-order Taylor approximation, preserving terms up to $\nu^{-4}$, as:
\begin{multline}
    \label{eq:refractive_index_expand}
    n \approx 1 - \frac{q^2}{8\pi^ 2m_e \epsilon_0} \cdot \frac{n_e}{\nu^2} \pm \frac{q^3}{16 \pi^3 m_e^2 \epsilon_0} \cdot \frac{n_e B \cos\theta}{\nu^3} \\- \frac{q^4}{128 \pi^4 m_e^2 \epsilon_0^2} 16\cdot \frac{n_e^2}{\nu^4} - \frac{q^4}{64 \pi^4 m_e^3 \epsilon_0} \cdot \frac{n_e B^2 (1+\cos^2\theta)}{\nu^4},
\end{multline}
\begin{figure*}
    \centering
    \includegraphics[scale=0.45]{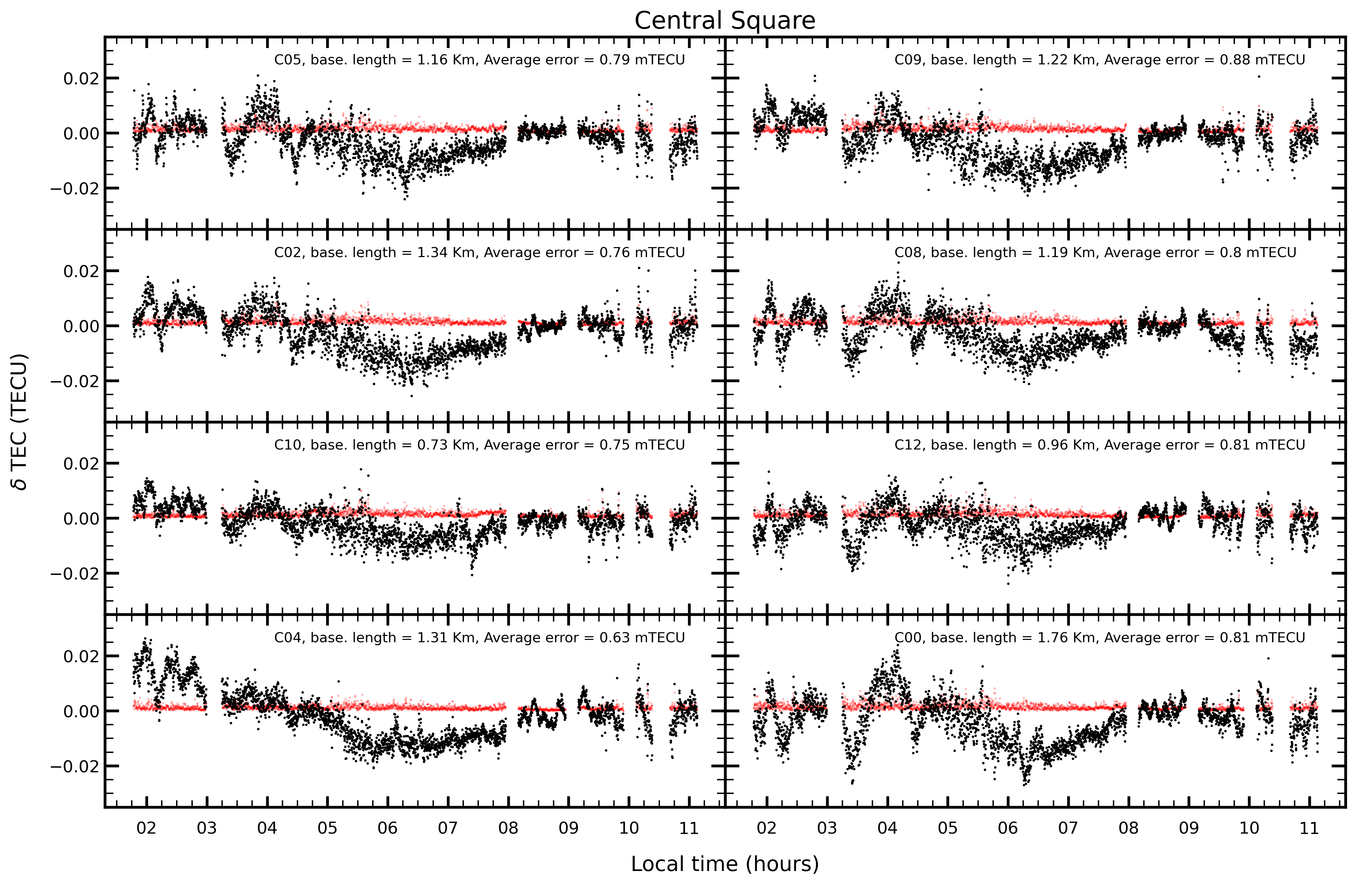}
    \caption{Total electron content variation (in TECU) along the observation. All values are differential between each of central square antenna and ``C13'' (chosen reference antenna). In each panel, the estimated uncertainty is also plotted in red.}
    \label{fig:centralsquare}
\end{figure*}
where, $q$ is the electron charge, $m_e$ is the electron mass, $\epsilon_0$ is the electric permittivity in vacuum, $n_e$ is the number density of free electrons, the magnetic field strength is denoted by $B$ and $\theta$ is the angle between $B$ and the propagation direction of an electromagnetic wave. The first term is the dominant term and is related to dispersive delay proportional to the TEC along the LoS. Higher terms can be ignored for frequencies greater than a few hundred MHz. The second term is due to Faraday rotation, where the positive and negative signs are associated with left-hand and right-hand polarised signals, respectively. The last two terms are usually ignored, but they are taken into account for observational frequencies below 40\,MHz \citep[see][]{Hoque2008RaSc...43.5008H}. \par
If ionospheric variations are not mitigated, the ionospheric phase terms can shift the apparent position of objects in the image plane. When higher terms take precedence, objects in the image plane become distorted and in extreme circumstances almost disappear. To accurately find astrophysical sources, calibration of the additional phase due to the ionosphere is vital. This phase term can help us study the ionosphere's dynamics. For calibration, the required time-dependent, direction-dependent, and antenna-based phase corrections can only be determined to sufficient accuracy using the technique of self-calibration or field-based calibrations. Additionally, there is the requirement of high time (order of 10\,sec) resolution calibration of the ionosphere to remove ionospheric-induced corruption from the interferometric data for observational frequencies below 300\,MHz. The distribution of calibrator sources across the field-of-view and a given array geometry leads to a dense sampling of the ionospheric electron density that is spatially variable. Compared with the sampling achieved from a dense grid of GNSS receivers dedicated to ionospheric studies over a local area, the radio-interferometric-based sampling is much finer. Thus, the possibility of synergy to study the ionosphere to remove corruption from interferometric observations is possible, which is not achievable solely by the dense grid of GNSS receivers over the globe \citep[see][]{Kassim2010amos.confE..59K}. \par
It is also crucial to note the higher sensitivity of a radio interferometer to detect any changes in the electron density or electron content in the ionosphere than the sensitivity provided by the present GNSS receivers. Ionospheric electron density fluctuations are affected by the nature of the solar disturbances, and the solar activity follows variabilities at different temporal scales. The variability of the dynamic ionosphere results from upward forcing by solar radiation. Various solar activities that include radio bursts show flicker noise characteristics as a function of time. Therefore, the presence of this corruption term in the measurement from a single antenna makes it necessary for performing high time resolution ionospheric calibration, making a strong requirement on signal-to-noise in order to perform fruitful ionospheric calibration. \par
\section{Exploring ionospheric structures using Antenna based Method}
\label{sec:antenna_based_method}
In this method to study the ionosphere, one observes a single bright source at the phase center and measures the varying phase on each baseline (a baseline refers to the distance between a pair of antennas) as a function of time. As electromagnetic signals from radio sources pass through the ionosphere, the first-order phase term, approximated using equations \ref{eq:phase_rotation} and \ref{eq:refractive_index_expand}, is given by:
\begin{eqnarray}
    \label{eq:phi2tec}
    \phi_{\rm ion} = 84.36 \left ( \frac{\nu}{\mbox{\scriptsize 100 MHz}} \right )^{-1} \left ( \frac{\mbox{\scriptsize TEC}}{\mbox{\scriptsize 1 TECU}} \right ) \mbox{ radians}
\end{eqnarray}
From equation \ref{eq:phi2tec}, one can infer that the difference in TEC along the LoS to the two antennas is directly proportional to the measured ionospheric induced phase. This method will be referred to as the \textit{antenna-based method}, which can study the temporal variation in differential TEC (\dtec) corresponding to the projected array onto the ionosphere. Several studies have used this method to probe the ionosphere.
\begin{figure*}[ht]
    \centering
    \includegraphics[width=2\columnwidth,height=10cm]{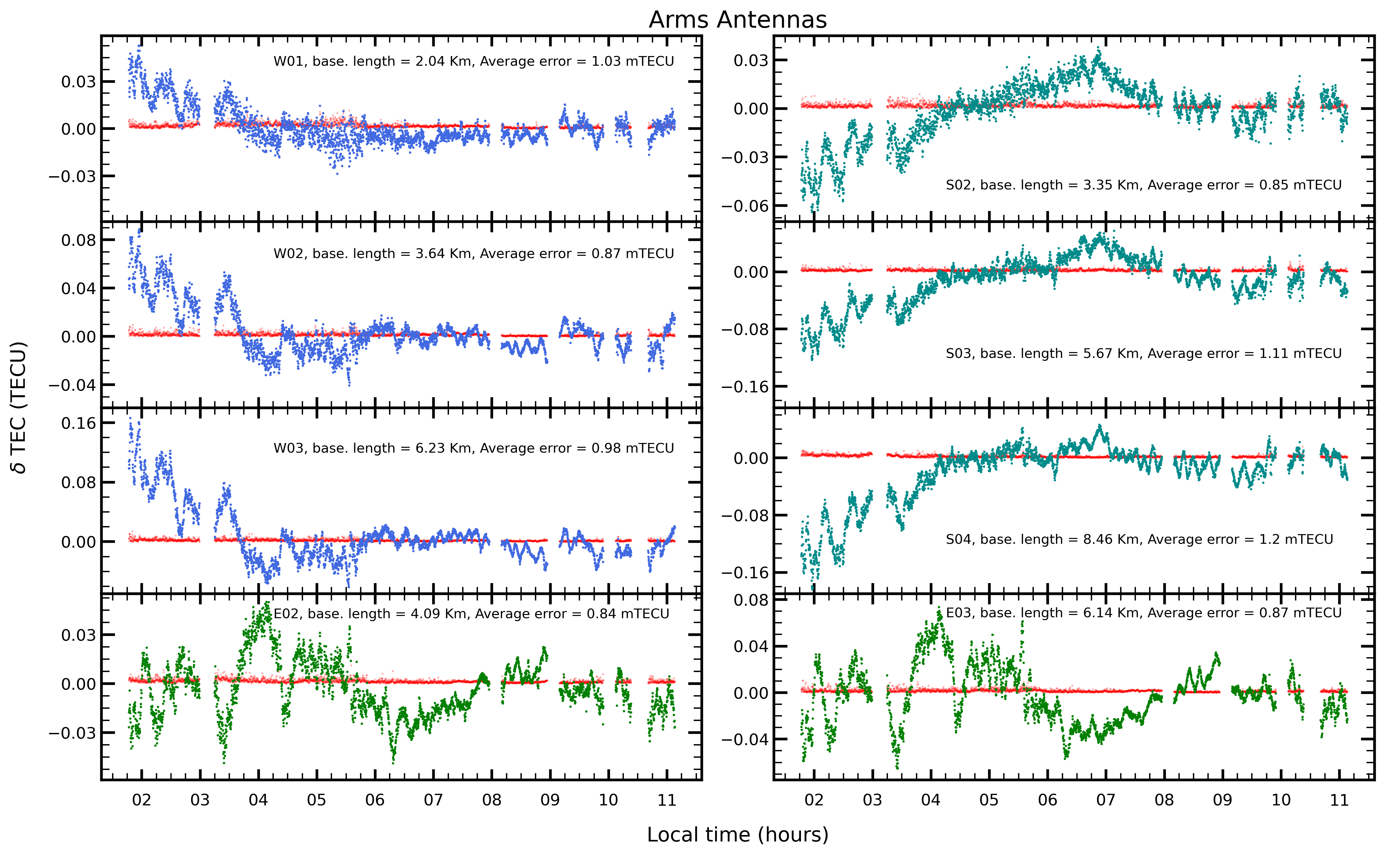}
    \caption{Same as Fig.\ \ref{fig:centralsquare} but along the arm antennas of the GMRT. Differential TEC (\dtec in TECU) along North-Western, Southern and North-Eastern antenna  are plotted in blue, teal and green respectively.}
    \label{fig:arms}
\end{figure*}
\citet{Helm2012RaSc...47.0K02H_temporal} using VLA observations of a single bright source in the FoV (Cygnus A) detected TEC variation with an amplitude of $< 10^{-3}$\,TECU and TEC gradients with an accuracy of about 0.2\,mTECU\,$\rm km^{-1}$. In the companion study, \citet{Helm2012RaSc...47.0L02H_spectral} conducted the spectral studies of TEC gradient measurements and could detect and classify many medium-scale traveling ionospheric disturbances (MSTIDs). Smaller waves were also detected which move in similar directions as the MSTIDs. \par
Using LOFAR, \citet{mev16} found that ionospheric irregularities (TEC) were anisotropic during the night with a precision of $10^{-3}$\,TEC when observing a bright quasar 3C\,196. A power-law behavior was also observed throughout the long-range baseline lengths using data solely from nighttime observations during the winters of 2012\,-\,2013, explaining the Kolmogorov turbulence in the ionosphere. Later, \citet{gasp18} demonstrated that scintillation corrupts the visibility amplitudes at ultra-low frequencies, resulting in an average of 30\% of the data loss during the night (compared to the daytime). Daytime observations are therefore encouraged, especially for the LOFAR-EoR experiment. Combined analyses of LOFAR, GNSS, and ionosondes \citep[see][]{Rich2020JSWSC..10...10F} reveal that if large-scale and small-scale TIDs are traveling perpendicular to each other, they cause instabilities that break down large-scale structures into smaller sizes. \par
We used GMRT observations obtained on August  06, 2012 from local midnight to post-sunrise hours pointed at an astronomical source, 3C\,68.2. The flagging and calibration of the dataset (at 235\,MHz) is performed using Common Astronomy Software Applications\footnote{CASA \url{https://casa.nrao.edu/}} (CASA), and the whole procedure is described in \citet{Mangla10.1093/mnras/stac942}. It is important to note that, during the observation, out of 30 antennas, four antennas were not operational and several time-stamps were flagged in nine antennas because of the low signal-to-noise ratio. Thus, this analysis was performed with the remaining 17 antennas (9 from central square and 8 from arms). Here, we will further study the outcome of the calibration (i.e. antenna-based complex gain solutions) when observing a bright radio source. The low-frequency observation with GMRT often has a low signal-to-noise ratio due to high sky temperature. To increase the signal-to-noise ratio, we used a longer time interval by averaging the data over several time-stamps while computing the complex gain solutions. As the ionosphere tends to vary very quickly, averaging over too many timestamps is not advisable. The calibration is done at a shorter time scale (10 seconds) to ensure accurate calibration and simultaneously track ionosphere changes. Note that the data was taken at a 0.5\,sec time interval. Thus, averaging over 20 time-stamps is a reasonable trade-off between decorrelation and signal-to-noise ratio. \par
The complex antenna gain solutions contain the effects of the ionospheric phase term along with instrumental noise. To mitigate the instrumental noise we follow a process called the continuum subtraction method \citep[see][]{Mangla10.1093/mnras/stac942,Helm2012RaSc...47.0K02H_temporal}. After obtaining the ionospheric phase term for each antenna, we converted it to differential TEC using the equation \ref{eq:phi2tec}. The resulting differential TEC (\dtec) is plotted for each antenna in Fig.\ \ref{fig:centralsquare} and \ref{fig:arms}. Furthermore, we estimated the uncertainty by calculating the median absolute deviation (MAD) of \dtec time series at each time-step for a particular antenna. When determining MAD for each time-step, the four closest time-steps are also taken into account to increase the accuracy of MAD computation. These MAD values are also plotted to demonstrate the relative accuracy by which \dtec is measured. The uncertainty in \dtec is of the order of $1\times10^{-3}$. The \dtec along each arm (Fig.\ \ref{fig:arms}) is scaled with baseline length (antenna with respect to reference antenna ``C13''), which shows that the \dtec is proportional to baseline length. Different patterns are observed along three arms (Fig.\ \ref{fig:arms}), thus the ionosphere not only varies in time but also in space. It also signifies that wave(s) are traveling in different directions or a dominant wave is propagating with smaller wave(s) (in different directions). By studying the spectral analysis of such patterns, one can extract the speed, direction and size of such dominant pattern(s) over the full array \citep[see][Mangla et al., in prep.]{Helm2012RaSc...47.0L02H_spectral}. \par
Measuring the TEC gradient is critical to understand TEC fluctuations better, as GMRT is sensitive towards \dtec between antenna pairs. Before computing the TEC gradient, geometric correction must be done to the data such that the measured \dtec (slant \dtec) would correspond to vertical \dtec using a thin shell approximation of the Earth's ionosphere located at an altitude of maximum electron density or ``peak height''. The peak height is estimated using the IRI Plas\footnote{IRI extended to Plasmasphere \url{http://www.ionolab.org/iriplasonline/}} software using the geophysical location of GMRT and time of the observation. It is to be noted that the IRI Plas shows the closest variation to the observed electron density or TEC values under disturbed conditions over the Indian longitude sector, compared to the other standard ionospheric models, due to IRI Plas model's ability to account for electron density distribution up to the plasmaspheric altitudes \citep[see][]{sc38}. The geometric correction is well described in \citet{Mangla10.1093/mnras/stac942} (Appendix A). After applying the correction to each antenna's \dtec time series, one can measure the TEC gradient over the entire array. \par 
\begin{figure}
    \centering
    \includegraphics[width=1\columnwidth,height=9cm]{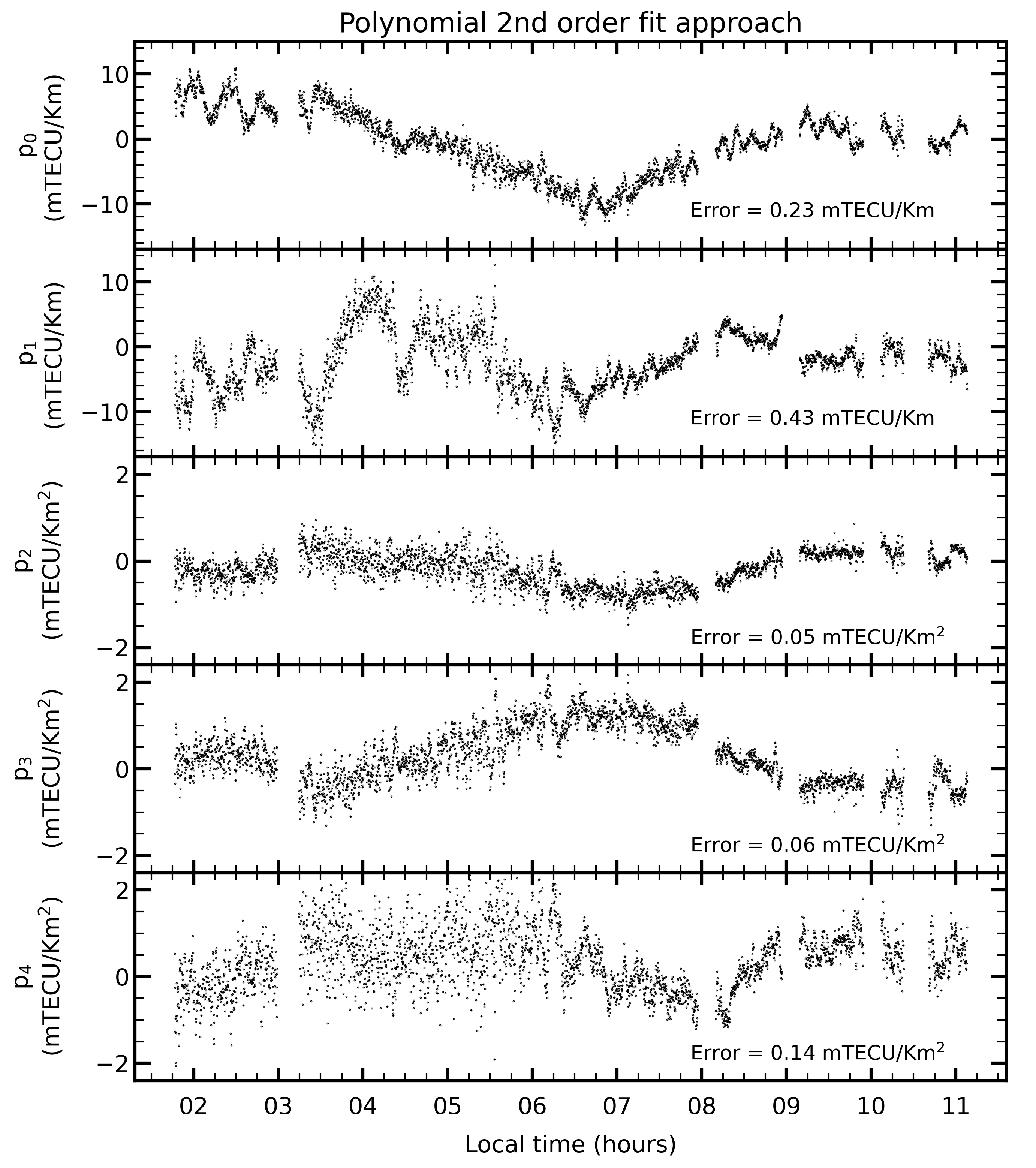}
    \caption{The fitted coefficients ($\rm p_0$ to $\rm p_4$) along the observation. These coefficients are fitted using the second-order polynomial equation independently for each time step using each of antenna pairs. The estimated-error for each coefficient is also mentioned in the respective panel.}
    \label{fig:polynomial}
\end{figure}
The two-dimensional TEC surface over the array can be computed using the second-order two-dimensional Taylor series (maximum baseline for the array is smaller than transient ionospheric waves), which has the following form:
\begin{equation}
    \label{eq:poly_2nd}
    \mbox{TEC} = p_{0}\,x + p_{1}\,y + p_{2}\,x^{2} + p_{3}\,y^{2} + p_{4}\,x\,y + p_{5}
\end{equation}
where x and y are antenna positions along with north-south and east-west directions. $\rm p_0$ to $\rm p_5$ are the polynomial coefficients. By calculating the difference of \dtec between antenna pairs (120 baselines) at each time step, one can increase the accuracy of these polynomial coefficients/parameters. Thus, equation \ref{eq:poly_2nd} transforms into the following form:
\begin{eqnarray}
    \label{eq:dpoly_2nd}
    \delta \mbox{TEC}_i - \delta \mbox{TEC}_j \! &=& \! p_0\,(x_i-x_j)+p_1\,(y_i-y_j)\nonumber \\ 
    &+& \! p_2\,(x_i^2-x_j^2) + p_3\,(y_i^2-y_j^2) \nonumber \\
    &+& \! p_4\,(x_i\,y_i-x_j\,y_j)
\end{eqnarray}
where subscripts $i$ and $j$ correspond to different antenna pairs (where $i\,\neq\,j$). It is important to note that the fitting is done independently at each time step, thus conserving the temporal and spatial TEC variation over the array. The obtained polynomial coefficients are shown in Fig.\ \ref{fig:polynomial} along with the standard error for each coefficient. One can easily notice that the amplitude for the $\rm p_1$ coefficient (along the east-west direction) is significantly high during the local nighttime. The same phenomenon is observed for $\rm p_0$ coefficient (along north-south direction) during the sunrise hour (around 6:00\,AM), which is a known behavior of MSTIDs, commonly detected around sunrise and sunset time \citep[see][]{her06}. Variations in higher-order coefficients ($\rm p_2$ to $\rm p_4$) are more significant during the nighttime, suggesting unanticipated ionospheric changes in and around the EIA region during these local times. \par
\section{Exploring ionospheric structures using Field Based Method}
\label{sec:field_based_method}
The wide-field radio interferometers can observe many celestial radio sources simultaneously, thus providing a large number of isotropically distributed pierce points \citep[see][]{Cohen_2009, Jordan2017MNRAS.471.3974J}. The phase is measured by analyzing the source position shifts rather than antenna-based solutions. This approach is known as the field-based method/calibration. The components of the TEC gradient along the LoS to the source averaged over the array and the time interval used to generate the image are proportional to the observed position shifts of a given source. As radio telescopes measure differential phases between pairs of antennas, the phase delay between the signal path arriving at a pair of antennas causes an apparent positional shift in the actual source location given by an angle:
\begin{equation}
    \label{eq:angle}
    \theta = \frac{\lambda}{L} \frac{\Delta\phi}{2\pi}
\end{equation}
where $\lambda$ is the observational wavelength, $L$ is the baseline length, and $\Delta\phi$ is the difference in ionospheric phase delays between pairs of antennas. If the ionospheric phase delay changes linearly across an array, all baselines ``observe'' the same position shift for a given source, and the image of a celestial source is merely moved from its true sky position. Using equations \ref{eq:phi2tec} and \ref{eq:angle}, TEC gradient along a direction $r$ on the ground to the observed angular shift in the same direction $\theta_r$, is given by \citep[see][]{Cohen_2009,Helm2020RaSc...5507106H}:
\begin{equation}
    \label{eq:position_shift}
    \frac{d}{dr} {\rm TEC} = 1.2 \times 10^{-4} \left (\frac{\nu}{\rm100 MHz} \right )^2 \left ( \frac{\theta_r}{1''} \right ) \rm TECU\,km^{-1}
\end{equation}
\begin{figure}[t]
    \centering
    \includegraphics[width=1\columnwidth,height=10cm]{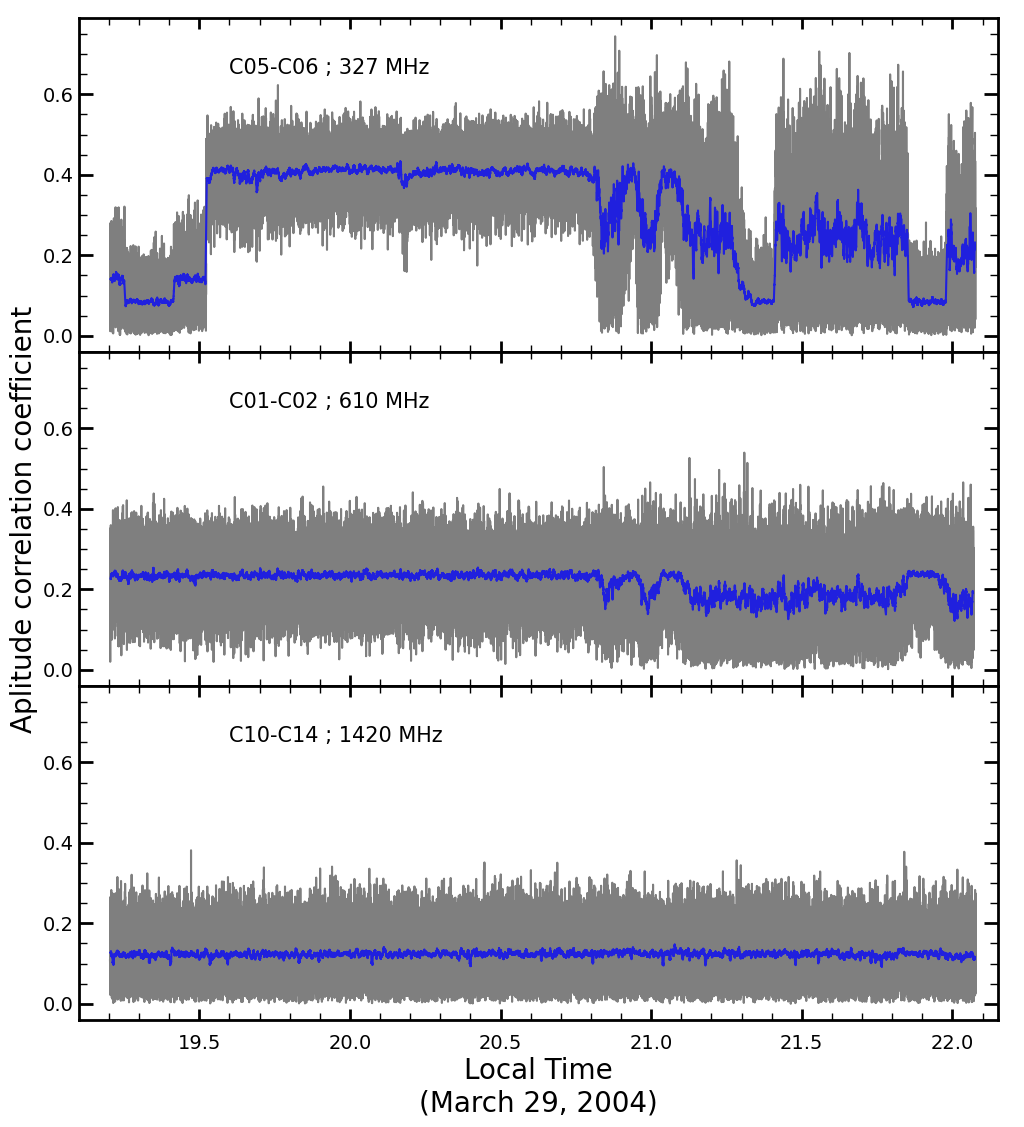}
    \caption{The 16\,sec moving averaged (blue curve) amplitude data and the corresponding amplitude (grey curve) of signal recorded at GMRT with a sampling interval of 108\,ms on March 29,\,2004 using signal from 3C\,218 at 327, 610, and 1420\,MHz respectively, during 19.2-22.1 Local Time (LT; UTC+05:30). [Adapted from \citet{ADas2008RaSc...43.5002D}]}
    \label{fig:amp_scintillation}
\end{figure}
\begin{figure}[t]
    \centering
    \includegraphics[width=1\columnwidth,height=10cm]{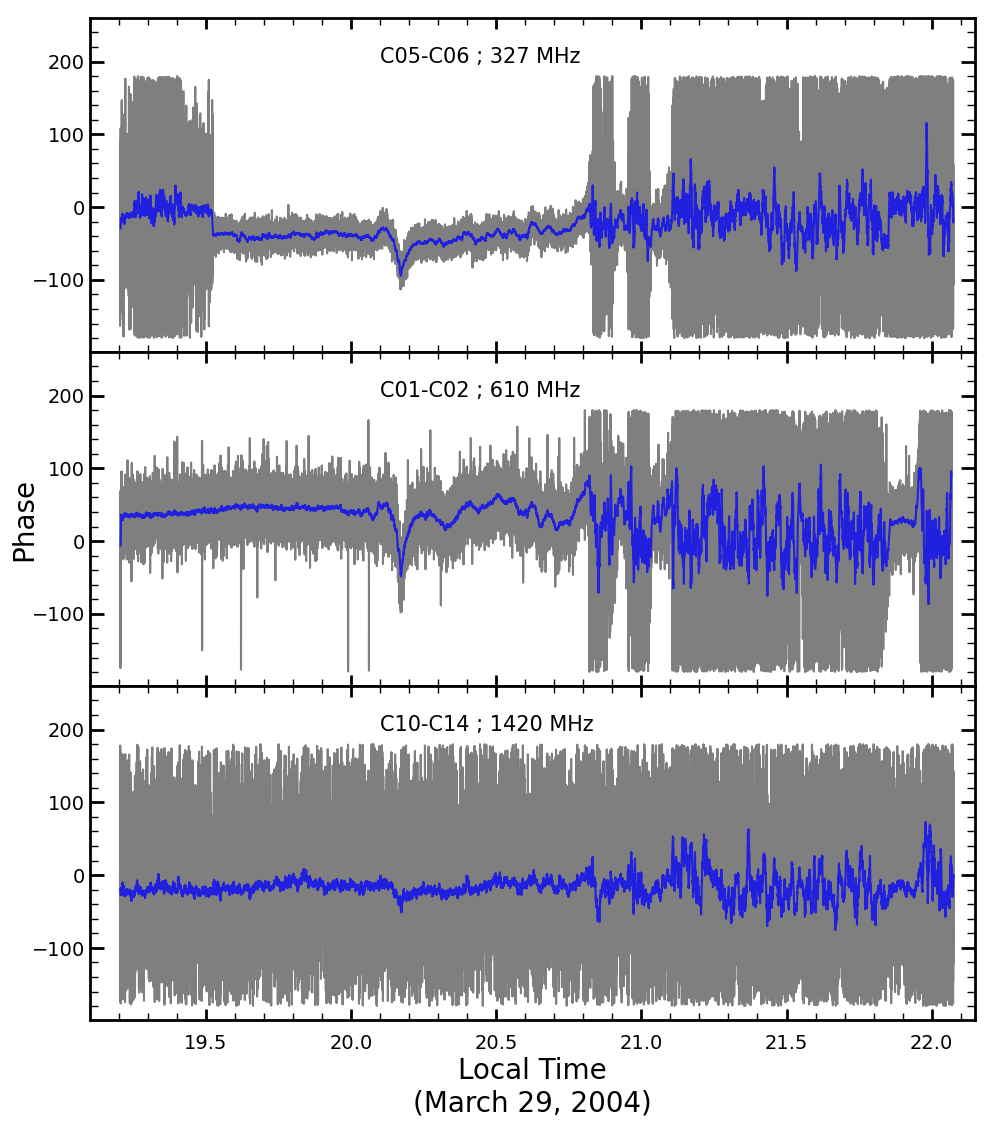}
    \caption{The 16\,sec moving averaged (blue curve) phase data and the corresponding phase (grey curve) of signal recorded at GMRT with a sampling interval of 108\,ms on March 29,\,2004 using signal from 3C\,218 at 327, 610, and 1420\,MHz respectively, during 19.2-22.1 Local Time (LT; UTC+05:30). [Adapted from \citet{ADas2008RaSc...43.5002D}]}
    \label{fig:phase_scintillation}
\end{figure}
If the ionospheric phase delay does not change linearly across the array and higher-order terms (curvature) are also present, the image will become distorted along with the actual sky position. \par
Several studies have used such methods to probe the ionosphere like \citet{Helmboldt2012RaSc...47.5008H} conducted the spectral analysis of 29 bright radio sources' positional shift in a single FoV to detect wavelike disturbances at 74\,MHz using VLA. Later, \citet{Loi2015RaSc...50..574L} investigated the ionosphere using a power spectrum study of ionospheric fluctuation (calculated using radio source positional offsets), which revealed TID characteristics. \par
\citet{Helm2020RaSc...5507106H} used 200\,hours of MWA's GLEAM survey \citep[see][]{Wayth2015PASA...32...25W} data to create images of ionospheric structures. The spectral analysis of these images revealed distinct characteristics in the nighttime ionospheric activity, like the generation of MSTIDs in association with sporadic E (Es). \par
To date, no field-based study has been accomplished using GMRT to the best of our knowledge. There are many observations of several deep fields carried out using uGMRT \citep[see][]{Arnab2019_2,Aishrila2020MNRAS.495.4071M} or all-sky radio survey (TGSS survey) at 150\,MHz \citet{TGSS2017A&A...598A..78I} which may be used to study the ionosphere over the Indian longitude sector.
\section{Estimation of Amplitude Scintillation using the GMRT}
\label{sec:scintillations}
The previous sections discuss about the ionospheric effect on the phase of the radio interferometers. In this section, we summarize a previous work by \citet{ADas2008RaSc...43.5002D} to show that the ionospheric effects on the amplitude of radio interferometer visibilities have also been detected in an earlier campaign in the declining phase of solar cycle 23. Using the GMRT, they have presented observational results related to the identification of the precursors of scintillations of the ionosphere on the TEC and the phase of a radio signal. The data was taken using different antenna pairs at multiple and broad range of frequencies (VHF to L-band) during March of 2004 which was the declining phase of $23^{rd}$ solar cycle. A representative case of March 29, 2004, has been shown here. They had recorded the amplitude and phase of the interferometric signal from 3C\,218 (a radio source) on March 29, 2004 at the frequencies 327, 610 and 1420\,MHz and at a sampling interval of 108\,ms. On the same day, they observed scintillation patches on the amplitude and the phase channels. Fig.\ \ref{fig:amp_scintillation} and \ref{fig:phase_scintillation} show these observed data during 19.20-22.10 LT for the GMRT antenna pairs C05-C06, C01-C02 and C10-C14 at 327, 610 and 1420\,MHz, respectively. \par 
It can be observed from Fig.\ \ref{fig:amp_scintillation} that there had been patches of scintillation that had occurred on March 29, 2004, in the 327\,MHz panel. The duration of these patches begin from 20.85-20.89 LT, 20.96-21.00 LT, 21.11-21.26 LT, 21.41-21.85 LT, and 22.00-22.08 LT. It is to be noted that on the same night, the signal's amplitude dropped down to zero during 19.20-19.52 LT, 20.87-20.90 LT, 20.95-21.02 LT, and 21.11-22.08 LT when the antenna was pointed away from the 3C\,218 radio source. Furthermore, from the 16\,sec moving averaged (blue curve) amplitude data, four distinct patches of scintillation appear as fading with respect to the background level from the observation at 327\,MHz and 610\,MHz frequencies. However, no such fading is observed at 1420\,MHz. Fig.\ \ref{fig:phase_scintillation} shows the 16\,sec moving averaged phase plots along with the phase data at 327, 610, and 1420\,MHz on this day, thereby illustrating the periodic nature of phase variation before the scintillation had occurred. From the panel showing the phase data recorded at 610\,MHz on this day, an interesting feature is observed in the form of periodic fluctuations before the scintillation onset. These quasi-periodic fluctuations or scintillation precursors occur during 20.10-20.25 LT in the phase records, followed by the first patch of scintillation that appeared at 20.85 LT. \par
\section{Forecast for SKA}
\label{sec:future_SKA}
In this section we have used the radio interferometric observations to estimate the ionospheric effects which is dependent on the differential TEC values.  Based on the observations and existing literature, we forecast the phase errors that SKA1-MID and SKA1-LOW will be susceptible to as a function of observing frequencies. We have also translated them into the possible dynamic range limit in imaging.
At ultra-low frequencies, differential Faraday rotation (second term in equation\ \ref{eq:refractive_index_expand} and higher-order terms) will also become essential and cannot be ignored. Using LOFAR, \citet{gasp18} showed that higher-order ionospheric effects are only prominent for observations below $\sim$40\,MHz with a maximum baseline of $\sim$50\,km. As SKA1-LOW observational frequency starts from 50\,MHz, higher-order terms (third-order onwards) can be ignored for core antennas, but their effects cannot be ignored for longer baselines. \par
By substituting  equation \ref{eq:refractive_index_expand} into equation \ref{eq:phase_rotation}, one can estimate the effects caused by first and second order terms \citep[see][Chapter 9]{Petit2010ITN....36....1P} associated with dispersive delay and Faraday rotation respectively:
\begin{multline}
    \delta\Phi_1 = -4840 \left( \frac{\nu}{100\ \rm MHz} \right)^{-1} \left( \frac{\rm d TEC}{1 \rm TECU} \right) [\rm deg] \\ 
    \delta\Phi_2 = \pm 38 \left( \frac{\nu}{100\ \rm MHz} \right)^{-2} \left[ \left( \frac{\rm d TEC}{1 \rm TECU} \right) \right. \\+ \left. \left( \frac{\rm TEC}{1 \rm TECU} \right) \cdot \left( \frac{{\rm d}B}{40\ \mu \rm T} \right) \right] [\rm deg];
\end{multline}
where we assumed magnetic field $B$\,=\,40\,$\mu$T with $\theta$\,=\,45$^\circ$. Under quiet-time ionospheric conditions, total TEC might change from $\sim$1\,TECU (during the nighttime) to $\sim$20\,TECU (during the daytime), affecting the second-order term. For example, consider a value for \dtec $\sim$0.3\,TECU, which is reasonable for a baseline of about $\sim$50\,km and observational frequency of 60\,MHz. Due to the first-order term, phase variations are generated several times of 2$\pi$. Second-order term or term due to Faraday rotation produces an effect of around $\pm$ 50$^\circ$/75$^\circ$ assuming d$B$=1\% at night/daytime. This effect will not be negligible for SKA1-LOW and needs to be addressed carefully (see Table \ref{tab:ska_low} for more detailed information). SKA1-MID observational starts from 350\,MHz, second-order will be negligible compared to the dominating dispersive delay (first-order). As the maximum baseline length for SKA1-MID is 150\,km, \dtec values will be around 1.5\,TECU, which is considerably large and needs to be corrected (see Table \ref{tab:ska_mid} for detailed information). \par
\begin{figure}[t]
    \centering
    \includegraphics[width=1\columnwidth]{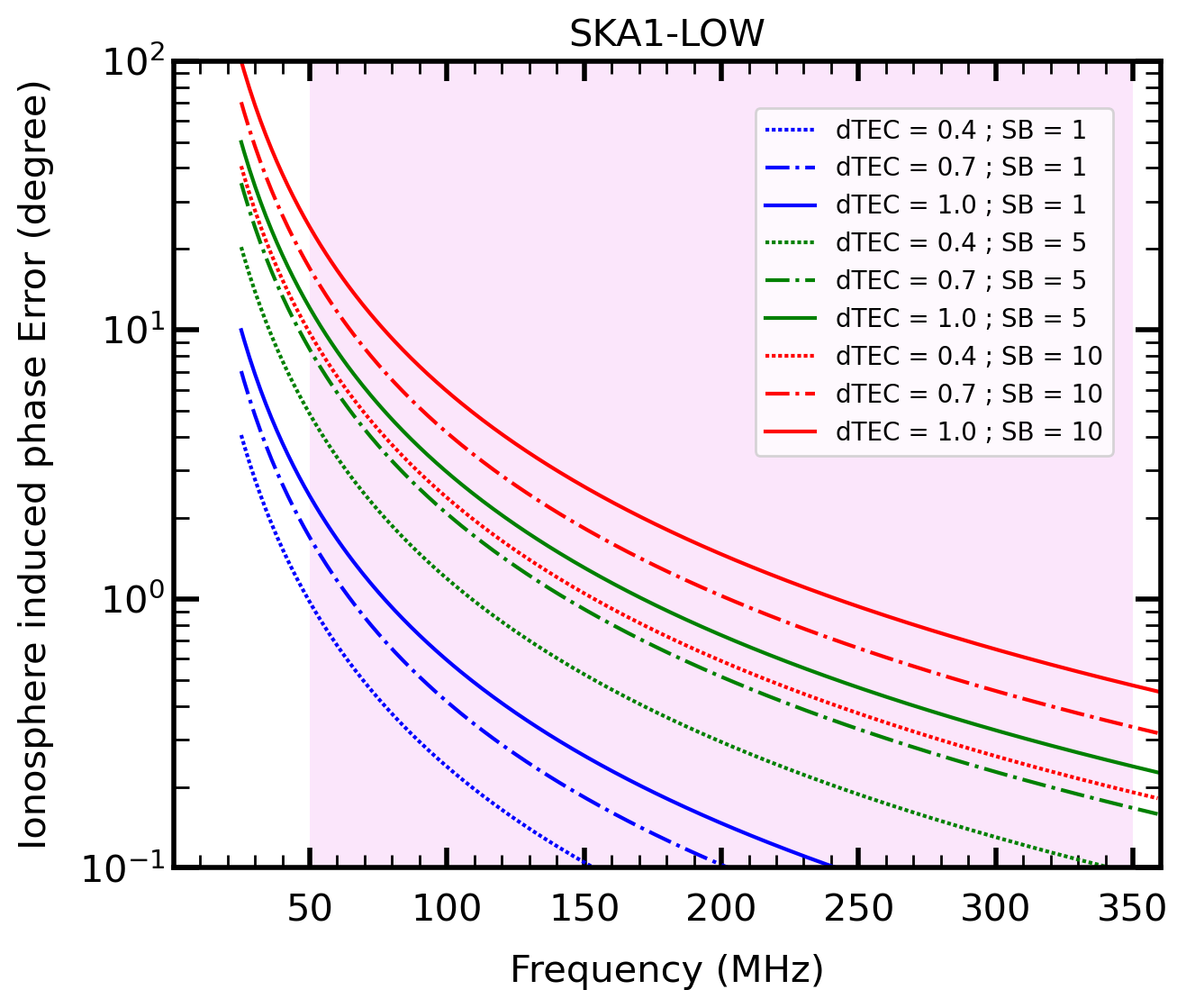}
    \caption{Ionospheric-induced phase variations between the beginning and end of a band of 1, 5 and 10 SKA1-LOW sub bands ($1 {\rm SB} \simeq 0.006$\,MHz). Assumed \dtec is 0.4, 0.7, 1.0\,TECU. These values were considered for baselines length of a few tens of kilometers. SKA1-LOW frequency bandwidth is 50-350\,MHz. (violet)}
    \label{fig:skalow_freqdependence}
\end{figure}
\begin{figure}[t]
    \centering
    \includegraphics[width=1\columnwidth]{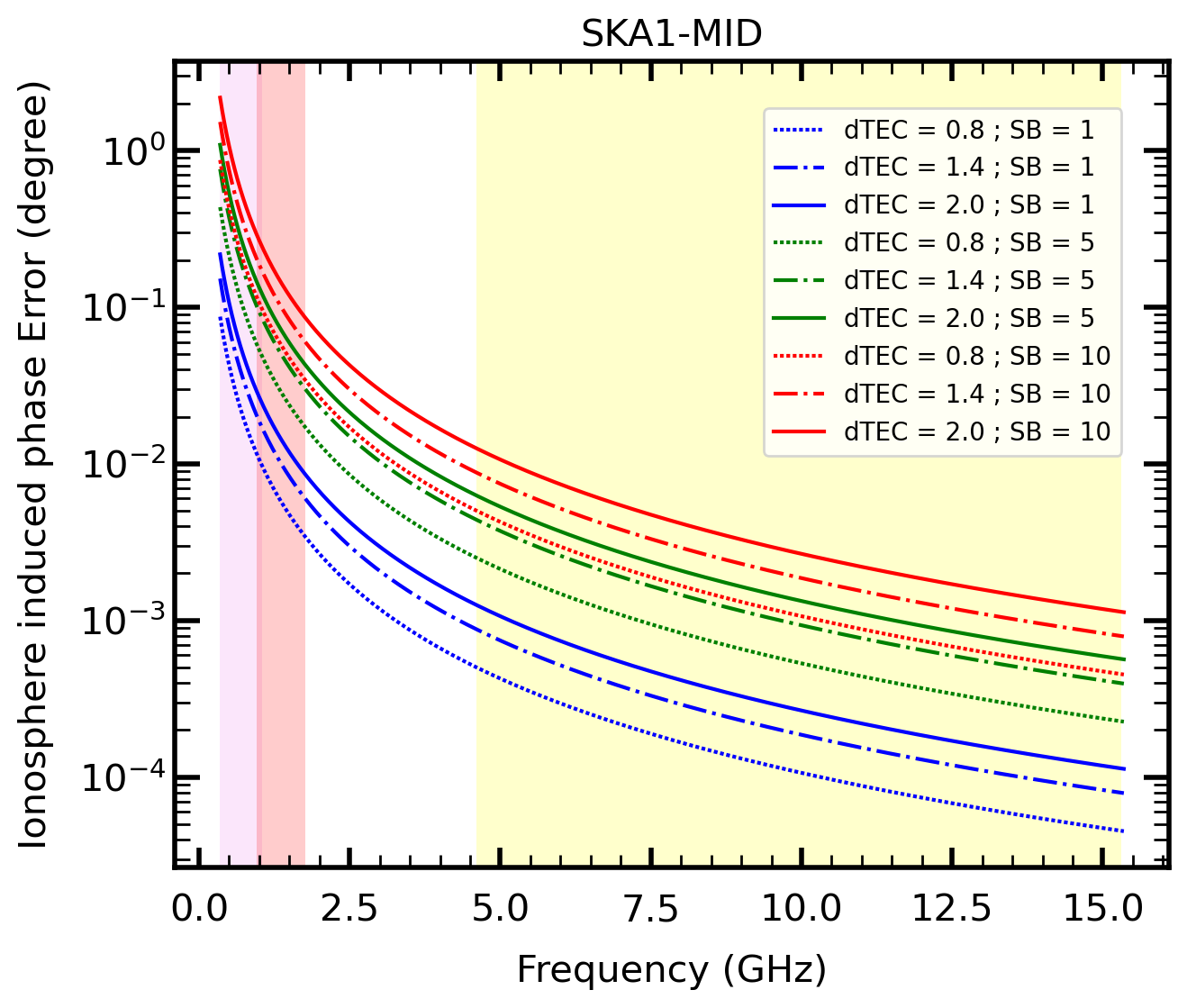}
    \caption{Ionospheric-induced phase variations between the beginning and end of a band of 1, 5 and 10 SKA1-MID sub bands ($1 {\rm SB} \simeq 0.01375$\,MHz). Assumed \dtec is 0.8, 1.4, 2.0 TECU. SKA1-MID frequency bandwidth are 0.350-1.05\,GHz (Band 1; violet), 0.95-1.76\,GHz (Band 2; red) and 4.6-15.3\,GHz (Band 5; yellow).}
    \label{fig:skamid_freqdependence}
\end{figure}

\begin{table*}[ht]
    \centering
    \caption{Typical ionospheric phase errors in degrees for SKA1-LOW}
    \begin{tabular}{lcccccc}
    \hline \\
                & I ord & II ord & I ord & II ord & I ord & II ord \\
    \dtec (TECU) &       & (day/night) &  & (day/night) &  & (day/night) \\
                & 50 MHz& 50 MHz & 100 MHz & 100 MHz & 250 MHz & 250 MHz \\ [1ex]
    \hline \hline \\ [0.1ex]
    1.0 (remote st., bad iono.) & 9680 & 181/153 & 4840 & 45/38 & 1936 & 7/6 \\
    0.4 (remote st., good iono.) & 3872 & 91/62 & 1936 & 23/15 & 774 & 4/2 \\
    0.05 (across FOV) & 484 & 38/9 & 242 & 9/2 & 97 & 2/$<$1 \\
    0.02 (Core St.) & 194 & 33/5 & 97 & 8/1 & 39 & 1/$<$1 \\ [1ex]
    \hline
    \end{tabular}
    \label{tab:ska_low}
\end{table*}
When observing a celestial radio source, most low-frequency telescopes work in low signal-to-noise ratio due to low antenna sensitivity and high sky temperature. A typical way to handle this is to average the data in longer time bins or frequency intervals when solving complex antenna gains. This is a trade-off between the signal-to-noise ratio and the decorrelation for finding a good solution. The ionosphere varies rapidly and averaging over more than 5-10\,sec time typically makes it difficult to trace these variations. Additionally, merging too many frequency channels is not a good idea, as shown in Fig.~\ref{fig:skalow_freqdependence}, between the edges of a five SKA1-LOW sub-band ($1 {\rm SB} \simeq 0.006$\,MHz) centered around 60\,MHz and considering \dtec value of 1.0\,TECU, there is a differential phase of 10$^\circ$. For core antennas (baselines less than a few kilometers, with \dtec $\lesssim0.4$\,TECU), this constraint can be relaxed, and one can average 10-20 sub-bands easily and still obtain good complex gain solutions. The same is shown for SKA1-MID ($1 {\rm SB} \simeq 0.01375$\,MHz) in Fig.~\ref{fig:skamid_freqdependence}, where averaging over high number of sub-bands will provide good complex antenna gain solutions as ionospheric induced phase errors are negligible at frequencies above 1\,GHz. \par
\begin{figure}[t]
    \centering
    \includegraphics[width=1\columnwidth]{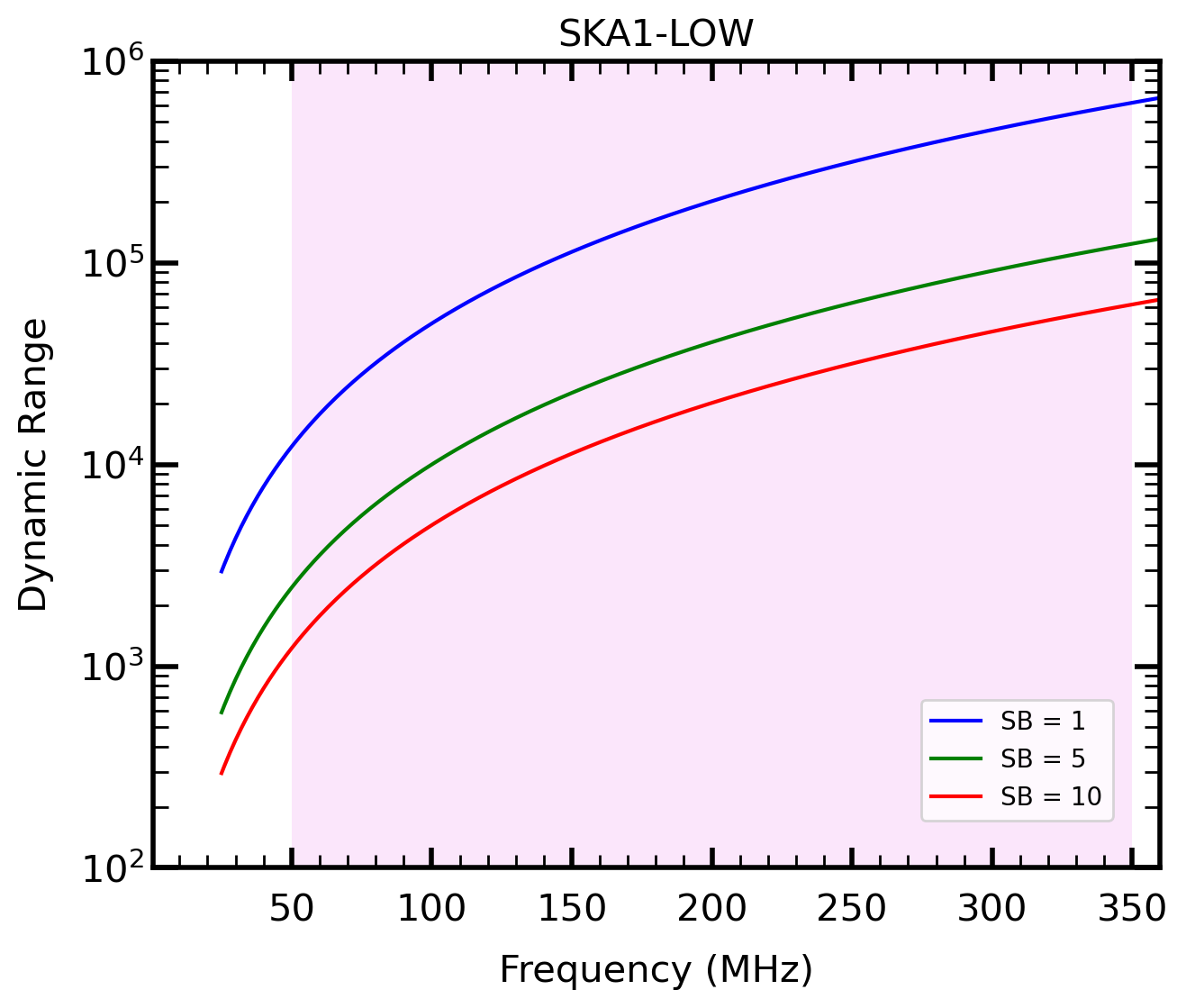}
    \caption{Dynamic range ($DR$) as a function of frequency for sub bands 1, 5 and 10 ($1 {\rm SB} \simeq 0.006$\,MHz). The phase error of these sub bands were estimated at \dtec = 0.7\,TECU. Number of station ($N$) is taken to be 512.}
    \label{fig:skalow_DR}
\end{figure}
In short, for accurate calibration for all baselines and different ionospheric conditions, antenna gain solutions must be calculated at high time and frequency resolution. Also, Figs.~\ref{fig:skalow_freqdependence} and \ref{fig:skamid_freqdependence} can be used to estimate the maximum of frequency channels averaging to compute the complex antenna solutions. \par
\begin{figure}[t]
    \centering
    \includegraphics[width=1\columnwidth]{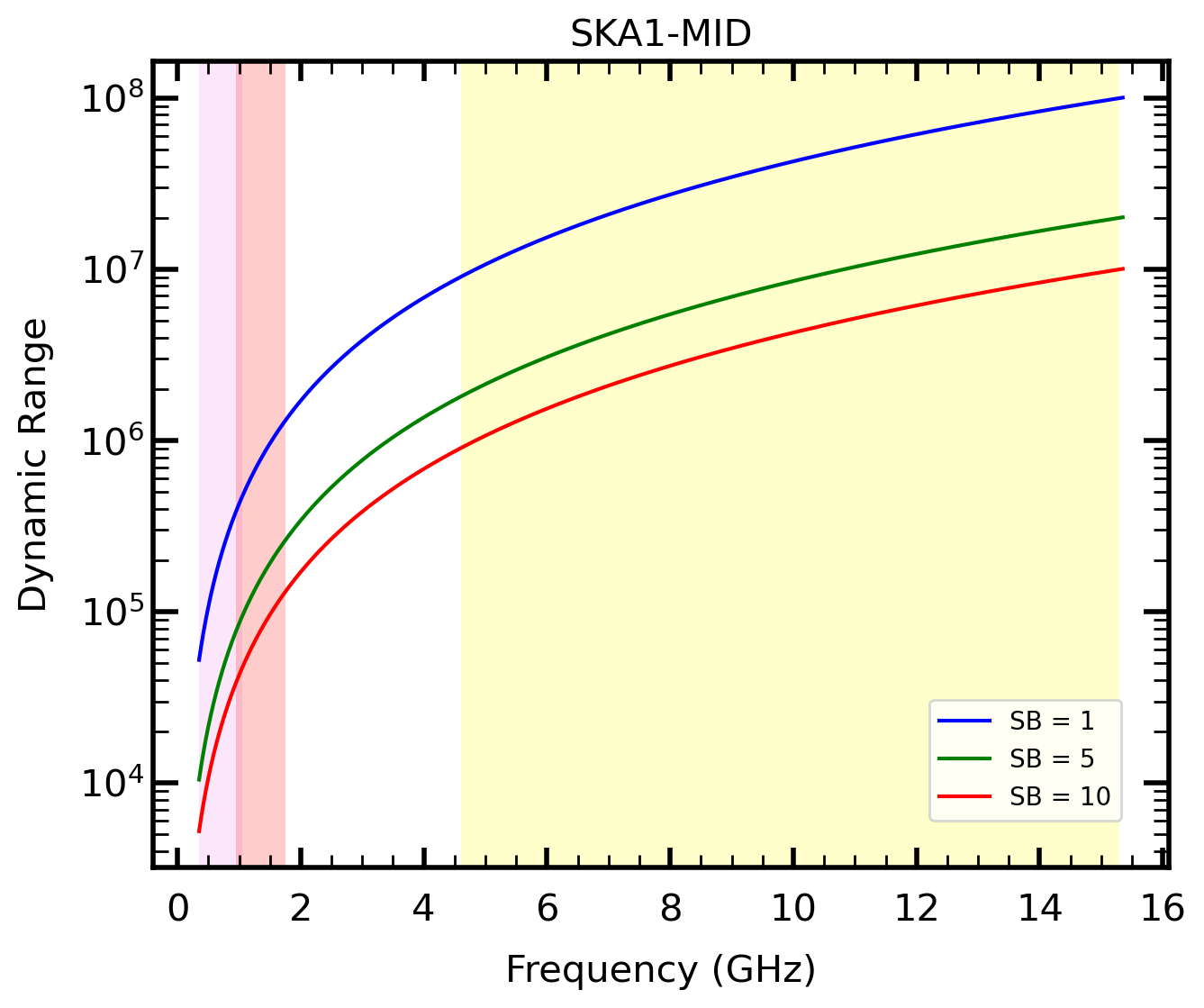}
    \caption{Dynamic range ($DR$) as a function of frequency for sub bands 1, 5 and 10 ($1 {\rm SB} \simeq 0.01375$\,MHz). The phase error of these sub bands were estimated at \dtec = 1.4\,TECU. Number of antennas ($N$) is taken to be 197.}
    \label{fig:skamid_DR}
\end{figure}

\citet{Perley1999ASPC..180..275P} estimated the limitation of dynamic range ($DR$) due to time-independent phase error ($\phi_{err}$ in radians) in all baselines, which will be introduced because of a point source data. The dynamic range of the image is given by the following:
\begin{equation}
    \label{eq:dynamic_range}
    DR \simeq \frac{1}{\phi_{err}}\sqrt{\frac{N(N-1)}{2}}
\end{equation}
\begin{table*}[ht]
    \centering
    \caption{Typical ionospheric phase errors in degrees for SKA1-MID}
    \begin{tabular}{lcccc}
    \hline \\
    \dtec (TECU) & I ord & I ord & I ord & I ord \\
                & 0.77 GHz & 1.4 GHz & 6.7 GHz & 12.5 GHz \\ [1ex]
    \hline \hline \\ [0.1ex]
    2.0 (remote st., bad iono.) & 1257 & 691 & 144 & 77 \\
    1.2 (remote st., good iono.) & 754 & 415 & 87 & 46 \\
    0.7 (across FOV) & 440 & 242 & 51 & 27 \\
    \hline
    \end{tabular}
    \label{tab:ska_mid}
\end{table*}
where $N$ is the number of interferometers antenna elements. We estimated the $DR$ for SKA1-LOW and SKA1-MID as a function of frequency for \dtec $\sim$ 0.7 and 1.4\,TECU, respectively, for sub bands (1, 5 and 10). From Fig. \ref{fig:skalow_DR} and \ref{fig:skamid_DR}, one can notice that $DR$ increases as a function of frequency, but at the same time averaging over too many sub bands will decrease it. It is important to note that, the $DR$ was estimated in view of a point source in the field. But, by considering the sensitivity of SKA1, the $DR$ estimates will be lower than mentioned in Fig. \ref{fig:skalow_DR} and \ref{fig:skamid_DR}, because many sources will be present in the FoV. \citet{Datta2009ApJ...703.1851D} showed using simulations that the phase calibration error of 0.1$^{\circ}$ for $N$ = 512 elements array at 158\,MHz will yield a $DR$ of $\sim\,10^{5}$ whereas the desired $DR$ is $\sim\,10^{8}$ to detect the \hi 21-cm signal from reionization.
\section{Summary}
\label{sec:summary}
One of the major factors limiting the quality of low-frequency radio interferometric observations is the effects of the ionosphere. The varying refractive index of the ionised plasma of ionosphere with time or space may cause dispersive delay, which can be directional-dependent. Indeed, it will affect the visibility phase measured by any interferometer. The GMRT's unique geophysical location can be used to better comprehend and characterise the ionosphere, which will aid radio astronomical studies at low frequencies observations \citep[see][]{Intema2017A&A...598A..78I, VanW2016ApJS..223....2V}. Some results using GMRT are as follows:
\begin{itemize}
    \item We have shown with the 235\,MHz data, that GMRT antenna-gain-based solutions lead to the precision of $1\times10^{-3}$ in \dtec measurements, which is an order of magnitude sensitivity higher in TEC measurements compared with GNSS-based measurements \citep[see][]{her06}.
    \item The two-dimensional TEC gradients over the array might be characterized using the polynomial-based techniques described in Sec. \ref{sec:antenna_based_method} This method efficiently recovers the features associated with larger amplitudes and longer periods of variation in a two-dimensional TEC gradient surface.
    \item In Sec. \ref{sec:scintillations}, the study by \citet{ADas2008RaSc...43.5002D} showed that visibility amplitudes and phases are affected by scintillation at frequencies 325 and 610\,MHz, when observing a bright radio source 3C\,218.
\end{itemize}
To improve the understanding of the ionosphere, field-based method detailed in Sec. \ref{sec:field_based_method} should be applied to the observation of several deep fields taken using uGMRT over the Indian longitude sector. \par
With the SKA construction phase underway, it is an excellent opportunity to develop calibration procedures and data analysis methodologies to detect ionospheric induced phase errors. Later, these methods may be utilised to mitigate Earth's ionospheric effects accurately from the measured visibility data, using low-frequency radio interferometers. 
\section*{Acknowledgements}
We thank the staff of the GMRT who have made these observations possible. GMRT is run by the National Centre for Radio Astrophysics of the Tata Institute of Fundamental Research. SM would like to thank the financial assistance from the University Grants Commission. SM further acknowledges Aishrila Mazumder for helpful discussions. The work of SC is supported by the Department of Space, Government of India. AD would like to acknowledge the support from CSIR through EMR-II No. 03(1461)/19. \par
This study also made use of \texttt{MATPLOTLIB} \citep[][]{matplotlib07} open-source plotting packages for \texttt{PYTHON}.

\bibliography{main}

\end{document}